Draft – Comments and Suggestions Welcomed
Version 2.0

Transforming the Structure of Network Interconnection and Transport

Douglas A. Galbi
Senior Economist[1]
Competitive Pricing Division
Common Carrier Bureau, FCC
January 10, 2000

Regulatory policy in telecommunications is desperately in need of mid-level theory. There is widespread consensus that regulators should try to promote competition. There is also bitter contention about what this implies for particular industry battles. For example, an industry observer declared, "Competition should be the policy. And code that enables competition should be the rule."[2] He then went on to argue that the Federal Communications Commission (FCC) should require AT&T to allow customers to choose the Internet service provider that provides service over AT&T's cable facilities. This reach to a policy action, without support from an analysis of the merits of different structural possibilities for competition, makes competition into merely a rhetorical device. In considering whether to mandate sub-loop unbundling, to require line-sharing, or to allow the use of unbundled network elements for the provision of leased lines, the issue is not just the costs of more aggressively promoting competition; such decisions also have profound implications for industry structure.

Regulators must make strategic choices among different structural possibilities for competition.[3] The common view that deregulation should occur once competition has developed is not a good framework for policy. Unlike antitrust policy, which acts to restore suppressed competition, pro-competitive regulation in a historically monopolized industry has to assess the merits of promoting new, different, and often incompatible dimensions of competition.[4] Waiting for competition to emerge before deregulating fosters the illusion that the existing structure of regulation does not shape the competition that develops. Of course it does, and the way that regulation shapes competition should be a key consideration in regulatory decisions. Regulators, while staying alert, humble, and flexible, need to think about this question: What are the most beneficial changes in industry structure that feasible regulatory policies could promote?

---

[1] The opinions and conclusions expressed in this paper are those of the author. They do not necessarily reflect the views of the Federal Communications Commission, its Commissioners, or any staff other than the author. I am grateful for numerous colleagues, at the FCC and in the global telecom policy community, who have shared their insights and experience with me.

[2] Lessig, Lawrence, "Cable Blackmail", The Standard, Nov. 14, 1999, on the web at www.thestandard.com/article/display/0,1151,5198,00.html.

[3] In considering the scope for strategic choice, there is an unfortunate tendency to gravitate toward poles of self-glorification ("we are, vis-à-vis the laws of nature in this new space, gods" [Lessig]) or of self-abnegation ("abolish the FCC" [Huber]). See Lessig, Lawrence, "Reading the Constitution in Cyberspace," 45 *Emory Law Journal* 3(1996) p. 7. Huber, Peter, *Law and Disorder in Cyberspace: Abolish the FCC and Let Common Law Rule the Telecosm* (New York: Oxford University Press, 1997), *passim*. For an insightful analysis of Huber's book, see Bell, Tom W., "The Common Law in Cyberspace" 97 *Michigan Law Review* (1999).

This paper attempts to expand discussion of this question. It puts forward three propositions:

**1. Structural problems are constraining beneficial developments for Internet services and voice telephony.**

With respect to Internet services, the challenges of defining meaningful products and establishing value-based interconnection relationships are slowing the development of new services requiring different billing protocols, qualities of service, and reliability. With respect to voice telephony, pro-competitive regulation that does not adequately consider the costs and benefits of promoting different forms of competition may perpetuate costly, complex regulatory battles and limit the scope for commercially driven business re-organization and service innovation.

**2. Development of competing, independently owned service interconnection points (SIPs) will stimulate development of local facilities and wide-area services.**

Connecting end users to a telecommunications network depends heavily on idiosyncratic, location-specific knowledge and equipment. In contrast, providing network services is inherently a non-local business relying on standardized routines and equipment. Businesses that I will refer to as service interconnection points (SIPs) would enable the separation of these two different spheres of activity and thus stimulate more dynamic, decentralized industry growth.

**3. Regulation of voice service interconnection should promote competing, independently owned SIPs.**

Existing regulatory authority and practice largely shape interconnection for voice telephony. While data traffic is growing much more rapidly than voice traffic, the value and bandwidth of voice traffic is still sufficient to influence strongly the over-all structure of network interconnection. Regulation should promote competing, independently owned SIPs by giving them the opportunity to have a privileged position for terminating voice calls.

---

[4] The goal of regulation in a formerly monopolized industry is not just to restrain the monopoly power of the incumbent. Cf. Lessig, "Cable Blackmail."

Taken together, these propositions suggest that feasible changes in the regulation of voice telephony interconnection could help to create a better competitive structure for the industry. Regulation thus far had tended to treat telephony interconnection architecture merely from the perspective of technical feasibility and particular competitors' requests. Moreover, telephony interconnection regulations have often been defined in terms of the incumbent operator's facilities and offices.[5] Given the enduring economic importance of interconnection architecture, a broader perspective should be considered.[6] The development of competing, independently owned SIPs would provide an industry structure conducive to better regulatory policy and more dynamic, decentralized industry growth.

**I. Structural limitations to the Internet**

A worrisome aspect of the current structure of Internet services is that retail customers often have little idea of what they are actually buying. A classic topic on Internet discussion lists might be stylized as: "I bought a T1 to the Internet. How do I make sure that I'm getting a full T1 to the whole Internet?" When buying leased lines, customers buy dedicated bandwidth between two specified points. Leased lines are well-defined, established products, but they have low average bandwidth utilization and high network management costs. When a customer purchases a T1 from an Internet service provider, the customer typically gets a "T1's worth" of dedicated bandwidth from the customer to the Internet service provider. But that connection is not in itself something that the customer values; the customer wants particular services with particular quality, reliability, and billing features.

A. Quality of service, reliability, billing: transactional, not technical challenges

Internet service providers currently face large challenges in attempting to provide their customers infrastructure-based service differentiation such as quality of service, reliability, and billing options. The unstructured and dynamic nature of interconnection on the Internet makes establishing new types of infrastructure-based services difficult. The long and tortuous discussions about "upgrading the Internet" from IPv4 to IPv6 illustrate the nature of the challenges. Lower profile examples of infrastructure development problems include small but annoying and persisting incompatibilities in e-mail formats, the handling of extended ASCII characters, and the treatment of e-mail attachments. Customers who, for a variety of reasons, need services not available using current, generic Internet connectivity can establish a variety of different forms of service level agreements, virtual private networks, customized peering arrangements, etc. Making such arrangements involves high transaction costs. That is a major weakness in the current structure of the Internet services industry.

---

[5] In the U.S., the Telecommunications Act of 1996 required incumbent local exchange carriers (incumbent LECs) to provide interconnection at any technically feasible point within their networks. Communications Act, amended, Section 251(c)(2)(B). To implement other requirements of the Act, the FCC has required incumbent LEC interconnection and network elements to be priced based on the forward-looking cost of the most efficient current network technology deployed within the structure of the incumbent LEC's wire centers (offices). *Implementation of the Local Competition Provisions in the Telecommunications Act of 1996, First Report and Order*, CC Docket 96-98 (released Aug. 8, 1996), para. 683-685. This approach, known as the FCC's TELRIC rules, makes interconnection rates and network elements prices dependent on the incumbent LEC's geographic structure of wire centers. The Commission of the European Communities has recommended maximum prices for local, single transit, and double transit interconnection, i.e. interconnection defined in terms of the hierarchy of the particular company's telephony network. See 98/195/EC: *Commission Recommendation of 8 January 1998 on interconnection in a liberalised telecommunications market* (Official Journal L 228, 15/08/1998 p. 0030-0034).

[6] The Peruvian regulator, OSIPTEL, has required the incumbent Peruvian operator, Telefonica del Peru, to provide at least one interconnection point in each of Peru's 24 departments. Thus OSIPTEL has made a choice about the geographic structure, but not the ownership structure or competitive structure, of interconnection points with Telefonica. See *Legislacion en Telecomunicaciones, Decreto Supremo No 020-98-MTC* para. 39 (available on the web at www.osiptel.gob.pe/marleg/cont/leg/leg/1998/ds20-98-mtc.htm).

Looking at the wholesale level (transactions between Internet service providers), some industry observers and participants have voiced concern that prevailing peering (interconnection) practices impede the Internet's development. One respected industry figure noted in early 1998, ". . .the extant non-policy peering policy [is] the biggest threat to the future of a competitive Internet."[7] A recent article in a leading trade publication declared:

> *Contrary to popular belief, the biggest impediment to a better, faster Internet isn't technological. It's political. The Internet is composed of about 8,000 smaller networks, and there are no rules (or laws) defining how they're connected. As a result, ISPs engage in lengthy, closed-door debates trying to determine how to connect, who should pay more, and how upgrades will be handled.*[8]

Lack of mutual understanding and acceptance of peering terms has led to disputes about traffic routing, traffic balances, and arrangements for international interconnection.[9]

These peering disputes point to the problem of trying to allocate service value rationally among networks interconnected without any established service structure. Suppose that network 1 has many customers that it charges for Internet access and network 2 hosts several servers that provide information over the Internet. Network 1 passes server queries to network 2, and network 2 returns the requested data to network 1. What do these two networks owe each other for network services? Network 2 sends many more packets into network 1 than it receives, so perhaps network 2 owes network 1 money for interconnection services. On the other hand, network 1's customers, who pay for Internet service only to network 1, requested the information from network 2, so perhaps network 1 should pay for receiving the information from network 2. Note also that the content providers who hire network 2 to host their content may have advertisers who want network 1's subscribers to see their advertisements, and they may also be collecting, from network 1's access customers, subscription fees for the content that they offer. Even this simplified market structure provides little guidance as to how to divide rationally service value between networks. Commercial negotiations over these issues are, not surprisingly, contentious.

---

[7] Rickard, Jack, "Editor's Notes," *Boardwatch* , May 1998 (on the web at boardwatch.internet.com/mag/98/may/bwm1.html, p. 6).
[8] Gareiss, Robin, "Tech Tutorial, Old Boys' Network," *Data Communications*, October 7, 1999.
[9] For a good discussion of Internet interconnection, see Huston, Geoff, "Interconnection, Peering and Settlements," available at http://www.isoc.org/inet99/proceedings/1e/1e_1.htm.

Two additional facts further confound attempts to allocate interconnection service value in an economically rational way. First, a particular service for a customer on one network may involve network traffic traversing two or more other networks, and the networks involved can change rapidly in response to changes in overall traffic patterns. Thus, for a particular service provided over the Internet, the networks that participate in supplying the service are not pre-defined. Second, the Internet is a platform for the provision of a wide variety of services, while interconnection between networks is typically not negotiated on a service-specific basis but instead assessed in terms of packet transport. The simple form that peering agreements currently take constrains transaction costs within an industry structure that offers meager guidance for allocating interconnection service value rationally. Such agreements, however, further obscure the economic signals relevant to interconnection service value.

The problem that Internet service providers have in defining differentiated infrastructure services for their retail customers is directly related to the problem of allocating interconnection service value rationally. From a supply perspective, the poorly developed and differentiated market for interconnection services constrains the development of retail products that use interconnection services as inputs. From a demand perspective, the economic transactions that define the current Internet industry structure generate only highly attenuated transmission of consumer value to agents making relevant network investment decisions. This industry structure provides a strong impetus to consolidation: services provided end-to-end within one company's network face none of the transactional challenges outlined above.

Moreover, without structural change in the Internet, small ISPs will make a much less beneficial contribution to the industry. An industry structure that gives small ISPs no opportunities to preserve and develop their businesses other than continually bargaining with a few large network operators will foster neither a dynamic industry nor decentralized innovation. Instead, small ISPs are likely to struggle to ensure their survival through political means. For insight into how this can occur, consider the position of the approximately 1400 small, independent local telephone companies in the U.S. These companies have effectively organized themselves through associations such as the National Exchange Carriers Association (NECA), the National Telephone Cooperative Association (NTCA), and the Organization for the Preservation and Advancement of Small Telephone Companies (OPASTCO). As it turns out, Sections 3(37)(B) and (C) of the U.S. Telecommunications Act include under the definition of rural telephone company all sufficiently small local telephone companies, irrespective of where they are located. Under the Act and associated FCC regulations, rural telephone companies are extended special regulatory and universal service benefits.

B. Public policy for Internet infrastructure: "don't change anything"

Public policy for Internet infrastructure has not responded to these increasingly serious industry challenges. The U.S. regulatory framework for Internet infrastructure was established fifteen to twenty years ago. The governing principle for Internet regulation is "don't change anything". An FCC order in 1980 decided that enhanced services are not subject to common carrier regulation.[10] Internet services have been classified as enhanced services, and hence they have not been subject to the many regulations that govern interconnection for voice telephony. Moreover, based on a 1983 FCC decision, enhanced service providers are treated as end users.[11] Thus Internet service providers can purchase flat-rated end-user offerings from local telephony companies and avoid a separate set of regulated prices, including per minute charges, that local telephony companies apply to switched voice telephony customers who are classified as "telecommunications carriers".

As argued above, the challenges associated with defining meaningful products and value-based interconnection relationships appear to be constraining the Internet from developing even more impressively than it is now. However, attempts to address these issues directly have not been successful. Consider, for example, the fate of the following significant initiative. In the summer of 1999, Reed Hundt, former Chairman of the FCC, attempted to set up an industry forum to address Internet interconnection. The forum's goal was described in the trade press as follows: "to stave off potential government regulation of peering by determining how large Internet service providers can fairly interconnect their networks with smaller counterparts."[12] The forum intended to address "financial settlements for interconnection and whether different charges should apply for different types of traffic."[13]

Several months later, a leading trade publication reported:
> "What I'm finding everywhere is indecision," he [Hundt] says. "There's a lot of interest, but no consensus." And that inability to find common ground is what concerns him. His original view was that peering would be settled in one of three ways: by a forum, market forces, or regulation. Now that the forum route seems not to be working, Hundt fears that regulators might step in.[14]

More than a year earlier, another leading trade publication noted, "Talk of any type of government intervention, from the Justice Department or the Federal Communications Commission, scares everyone in the game."[15] One might imagine that having a former Chairman of the FCC discuss these issues would heighten concerns. The failure of this effort and others[16] to make progress suggests the great difficulty of trying to resolve Internet interconnection issues directly given current industry structure.

---

[10] *In the Matter of Amendment of Section 64.702 of the Commission's Rules and Regulations (Second Computer Inquiry)*, 77 FCC 2d 384, 419 (1980) (Computer II Final Decision).
[11] *MTS and WATS Market Structure*, 97 FCC 2d 682, 711-22 (1983).
[12] "Hundt to Set Up Peering Forum", *Data Communications* (June 1999), on the web at http://www.data.com/story/DCM19990610S0001.
[13] Ibid.
[14] Gareiss, Robin, "Tech Tutorial, Old Boys' Network", *Data Communications* (October 7, 1999).
[15] McCarthy, Bill, "ISPs Agree on Little, But They Don't Want the Government to Choose Their Peers," Miscellaneous in *Boardwatch* (May 1998).
[16] See, for example, "Brokered Private Peering (BPP)tm Group", Miscellaneous, *Boardwatch* (May 1998).

## II. Telephony regulation: promoting competition without judgement

Policy makers around the world, of many different political persuasions, now regularly proclaim their determination to promote competition in telecommunications. Adding substance to this determination requires addressing the following question: "For what should competition be promoted?" Promoting competition for particular services can have major implications for the evolution of regulation and the long-term competitive structure of the industry. Unfortunately, the "competition for what?" question has not received adequate consideration. This policy deficiency may have significant long-run costs: current voice telephony regulation may be unintentionally promoting an industry structure for competition that is much less beneficial than alternatives.

A. Avoiding judgements about competitive structure

Despite the enactment of a broad range of pro-competitive policies in telecommunications, policy analysts and policy makers are reluctant to confront directly the question of competitive structure. Instead, policy makers are assumed to be responsible for promoting competition for everything, everywhere. Seldom today is there any consideration of whether this is feasible or desirable, although regulators are also generally expected to pledge to protect consumers from any unfortunate effects of all this competition.[17] "Competitive neutrality" is also considered to be a key policy concept. It appears to mean that the regulator, while promoting competition for everything, will ensure that regulation doesn't promote one type of competition more than another. Such rhetoric, while incoherent, emphasizes that regulatory considerations of competitive structure are preparations for exercising regulatory discretion. Regulatory discretion is considered an undesirable and unnecessary aspect of sound pro-competitive policy.

1. Failure of the technocratic approach

In traditional neo-classical economics and public utility theory, analyses of technology and demand provided the analytical basis for judgements about competitive structure. The key phrase in this approach is "natural monopoly". Based on estimates of the characteristics of production functions and consumer demand, the industry is partitioned into markets, and the markets are classified as either "workably competitive" or "naturally monopolistic". Regulators promote competition in markets that are "workably competitive" and continue to regulate markets that are " naturally monopolistic".[18]

---

[17] For some analysis of whether more competition in U.S. long distance services is an appropriate policy goal, see Galbi, Douglas, "the price of telecom competition: counting the cost of advertising and promotion," *info* vol 1, no. 2 (april 1999) pp. 133-193, and Galbi, Douglas, "Regulating Prices for Shifting Between Service Providers," currently available through the Social Science Research Network (www.ssrn.com), forthcoming in *Information Economics and Policy*.
[18] Tim Brennan provided me with comments that have formed the substance of this paragraph.

Unfortunately, the technocratic approach provides limited guidance for pro-competitive policy in the telecommunications industry of today. When telephony was a radically different business from cable service, when switching cost weren't driven by developments in the computer industry, and when wireless telephony and the Internet didn't exist, economists could and did analyze whether particular parts of the telephone network were a natural monopoly. Today's much more dynamic technological and demand context make such analysis much more difficult.

More significantly, this technocratic approach obscures issues that should be central policy considerations. First, the effects of regulation on the structure of competition should be a key regulatory concern. Interconnection regulation, for example, might be able to mitigate economies of scope that would otherwise preserve a "natural monopoly". More generally, the classification of "workably competitive" markets cannot be prior to consideration of regulation, which is no less natural than markets. Second, time should be a key regulatory concern. Facilities-based broadband connectivity to residences, for example, might develop into a "workably competitive" market over a time span that is too long to deliver benefits comparable to a loop unbundling requirement imposed on the incumbent telephony provider.

The technocratic approach largely fails to inform current telecommunications policy. A determination to promote competition everywhere in the telecommunications industry might be taken to imply the judgement that there are no natural monopolies in the industry. But policy makers have not made that judgement. Policy makers now generally ignore the technocratic approach to making judgements about competitive structure, not because of its weaknesses, but because they now largely attempt to avoid making judgements about competitive structure.

2. Focus on incumbent monopolists' products and networks

One way in which pro-competitive policy attempts to avoid making judgements about competitive structure is by assuming that the incumbent monopolists' products and network elements define the relevant competitive possibilities. Crude forms of this perspective include the view that competition will lower prices but will not change anything else, or that emerging wireless communications services will evolve to be like the services that wireline local exchange carriers provide.[19] More sophisticated examples of this policy perspective include the FCC's attempt to establish competition for switched voice transport to incumbent end-offices.[20] More recent policies focusing on more extensive forms of unbundling of the incumbents' networks implicitly tend to sanction the view that competition, if it develops, will develop for elements of the incumbents' networks. One sees a slightly twisted version of this perspective in a recent paper that implicitly proposes that unbundling should not be imposed to enable competitors to provide products that the incumbent operator does not provide.[21]

---

[19] For a critique of the latter view, see Charles D. Cosson, "You Say You Want A Revolution? Fact and Fiction Regarding Broadband CMRS and Local Competition," CommLaw Conspectus 7:2 (Summer 1999) pp. 223-278.
[20] A detailed description and analysis of this policy was part of an earlier draft of this paper, available on request from the author.

3. The rise of the client-driven approach

Another way regulators avoid making judgements about competitive structure is by taking a client-driven approach to pro-competitive regulatory policy. This approach works as follows. Company B argues that the regulator must require Company A to do X so that Company B can provide service Y. Typically Company A is described as a monopolist and X is described as essential for providing Y. Moreover, Y is described as a service that many customers want, because it will be better or lower priced than the alternative. The regulator promotes competition by requiring Company A to do X. The regulator does not require Company B to actually provide Y, nor does the regulator require Company B to provide Y better or lower-priced than the identified alternative; such requirements would be considered intrusive regulation. What sorts of competition are promoted thus depend on companies' requests, the regulator's responsiveness to them, and what companies actually do with the regulatory rights that they gain.

The U.S. Telecommunications Act of 1996 supports this approach, but it does not require it. Section 251(c)(3) of the Act identifies duties of incumbent local exchange carriers to provide access to network elements in response to requests from telecommunications carriers seeking to provide telecommunications services. Section 251(d)(2) gives the FCC the responsibility for defining the scope of these duties. Under Section 251(d)(2)(B), an "impair" limitation on access duties is set out in the context of a telecommunications carrier "seeking access to provide the services that it seeks to offer". However, the "necessary" limitation on access duties in Section 251(d)(2)(A) is not put in this context. One might imagine a form of access that is necessary to provide a telecommunications service that a telecommunications carrier seeks to offer, but that is not necessary to promote a pro-competitive, deregulatory strategy for the industry. Moreover, Section 251(d)(2) sets out minimum access standards that the FCC must consider. Additional access standards associated with a broader vision for pro-competitive, deregulatory industry development appear to be permissible, but they have not been explored.

B. Unintentional aspects of emerging industry structure

Although policy makers have been pre-occupied with the extent of competition and its speed of development, industry performance also depends on other aspects of competitive structure. Even in a communications industry in which all product markets are workably competitive, at least two sorts of potential weaknesses could exist. First, the industry, although competitive, might not be capable of re-organizing itself quickly to adjust to changes in technology or the scope of potential trades. Second, the industry, although competitive, might dissipate significant economic value as companies continually appeal to the regulator to decide narrow, complex issues concerning the distribution of value between companies. Aspects of pro-competitive regulatory policies for voice telephony may be contributing to the development of such weaknesses.

---

[21] See condition (5) in the unbundling standard proposed in Jerry A. Hausman and J. Gregory Sidak, "A Consumer-Welfare Approach to the Mandatory Unbundling of Telecommunications Networks," The Yale Law Journal, vol. 109 pp. 418-505.

Current pro-competitive regulatory policies for voice telephony are increasing the cost of adjusting a geographic structure of incumbent end offices that is probably highly inefficient. The geographic structure of incumbent telephone operators' end offices was largely established prior to 1917. Given subsequent dramatic developments in switching and transport technology, this structure is likely to be highly inefficient. Pro-competitive regulation, however, is now deeply connected to the existing structure of incumbents' networks. Such regulations have partitioned incumbents' networks into elements defined in terms of incumbents' existing end-offices. Regulated rates for interconnection have been defined in terms of tandem and end-office hierarchies. Competitors have granted regulatory rights to collocate in incumbent end offices. Given such regulations, changes in the structure of incumbents' network are likely to occur much more slowly.

Moreover, collocation rules have created narrow, complex, and enduring regulatory battles between companies. State regulatory commissions in the U.S. began to require collocation in incumbents' offices as early as 1989.[22] The FCC began to establish national collocation rules in 1992.[23] Yet companies are still battling intensely over narrow issues that affect the value of collocation obligations. For example, a 1999 FCC order decided, among other issues, that collocating carriers are allowed to construct their own facilities for cross-connecting among themselves, that incumbent LECs must provide "shared collocation", "cageless collocation", and collocation in adjacent controlled environments if collocation space is exhausted.[24] The FCC also decided:

> *. . .an incumbent LEC that denies collocation of a competitor's equipment, citing safety concerns, must provide to the competitive LEC within five business days a list of all equipment that the incumbent LEC locates within the premises in question, together with an affidavit attesting that all of that equipment meets or exceeds the safety standard that the incumbent LEC contends the competitor's equipment fails to meet.[25]*

A large number of other issues, at this level of detail, are emerging with respect to loop-sharing, managing interference among loops ("cross talk"), sub-loop unbundling, and the use of various combinations of wholesale network services or elements.

---

[22] The New York Public Service Commission (NY PSC) required virtual collocation for private line service in 1989, and physical collocation for switched services in 1992. For more details and citations to NY PSC orders, see *Application by Bell Atlantic New York for Authorization Under Section 271 of the Communications Act to Provide In-Region, InterLATA Service in the State of New York*, FCC Memorandum Opinion and Order, CC Docket 99-295, FCC 99-404 (rel. Dec. 22, 1999).

[23] Initially, the FCC required physical collocation. This requirement was overturned in court, so the FCC enacted a virtual collocation requirement. The Telecommunications Act of 1996 give the FCC authority to require physical collocation.

[24] *Deployment of Wireline Service Offering Advanced Telecommunications Capability*, 144 FCC Rd 4761, CC Docket 98-147 (rel. Mar. 31, 1999).

[25] Ibid para. 36.

Attempts to implement unbundled access to the incumbents' network have quickly led to the recognition that information systems for pre-ordering, ordering, provisioning, repair, maintenance, and billing (OSS) determine the usefulness of physical facilities. Regulators are thus forced to confront complicated issues associated with information systems' capabilities and performance. To get a sense for the nature of some of the regulatory fights between parties, consider that parties expressed concern, and the FCC noted concern, that competing carriers using Bell South's network must scroll through lists of products and services to fulfill particular customer orders. In contrast, Bell South's own retail interface allows its customer service representatives to find a product or service simply by typing the first few letters of the product's name.[26] A recent independent test of Bell Atlantic's OSS in New York State, done for regulatory purposes, involved statistics for 855 test elements.[27] Such tests merely establish performance. Fertile opportunities still exist for disputing blame in a process somewhat similar to trying to figure out why your operating system crashed when you tried to run a particular feature of a particular application program.

Detailed regulations and adjudications related to network operators' information systems are likely to have high welfare costs. Technology is rapidly driving down the cost of switching and transport hardware.[28] The most important challenge in improving industry performance is not to promote the more efficient deployment of network hardware. It is to foster a wider range of network management capabilities and to promote quicker, more customized service.[29] Regulations that require an incumbent network operator to provide to competitors non-discriminatory access to its OSS greatly diminish the incumbent's incentives to improve its OSS.[30] Subjecting the performance of such systems to an adversarial process also makes change more risky and more difficult.

In regulated industries, companies and customers tend to acquire quasi-property rights in existing arrangements. As companies and customers make plans and investments based on the wide range of regulatory rights being established for telephony, the possibilities for future regulatory dis-engagement narrow, and the regulator may be required to police continually the value of acquired rights. One might hope that emerging market opportunities would gradually erode the value of rights acquired under telephony regulation. For example, suppose that under telephony regulations a company has won the right to do "x" for price "p1". If competitive industry development results in "x" being freely offered for price p2<p1, then regulatory rights will be liquidated in a decentralized way. Regulatory rights will also be liquidated if a more cost effective opportunity than "x" emerges, or if for some other reason "x" is no longer a useful right.

---

[26] *Application of BellSouth Corporation, et. al. Pursuant to Section 271 of the Communications Act of 1934, as amended, To Provide In-Region, InterLATA Services in South Carolina*, 13 FCC Rcd 539, CC Docket 97-208 (rel. Dec. 24, 1997), para. 174.

[27] *Bell Atlantic NY 271 Application*, Introduction and Summary.

[28] In 1998, software accounted for 40% of capital spending for MCI WorldCom, with switching and transport accounted for 6% and 19% respectively. In contrast, in 1988 software accounted for 3% of MCI's capital spending, while switching and transported accounted for 19% and 44% respectively. MCI WorldCom expects that switching and transport will amount to less than 25% of capital spending per year within a few years. John Sidgmore, MCI WorldCom Vice Chairman, presentation to stock analysts, Spring 1999.

[29] In 1998, MCI WorldCom's cost for delivering one minute of voice traffic broke down as follows: OSS, 5%; switching, 3%; transport, 4%; operating costs, 26%; access (interconnection fees), 62%. John Sidgmore, MCI WorldCom Vice Chairman, presentation to stock analysts, Spring 1999.

[30] Under the Section 271 requirements of the U.S. Telecommunications Act, as implemented by the FCC, "For OSS functions that are analogous to those that a BOC provides to itself, its customers, or its affiliates, the nondiscrimination standard requires the BOC to offer requesting carriers access that is equivalent in terms of quality, accuracy, and timeliness. The BOC must provide access that permits competing carriers to perform these functions in "substantially the same time and manner" as the BOC. *Application by Bell Atlantic-NY for Authorization Under Section 271*, CC Docket 99-295, para. 85. Such regulation lessens the competitive value to the RBOC of investing to improve its OSS.

The liquidation of regulatory rights is not, however, a result that will follow inevitably from industry growth and the development of competition. For example, if incumbent LECs' end offices become key network interconnection points, then collocation rights may endure long beyond the growth of competitive telecommunications networks. Regulated rights associated with complex information systems like OSS are likely to be difficult to liquidate because of idiosyncratic investments that companies will have to make to establish compatible software and necessary business routines. Policy makers and industry participants who look forward to the growth of competition and deregulation need to consider the possible significance of what economists call "hysteresis" or "lock-in" effects. One can easily image the rapid growth of "grandfather" clauses that perpetuate the effects of legacy regulations long beyond the relevance of the policy concerns that motivated them. Competition will occur, but in the context of regulations that hinder industry change and foster wasteful regulatory battles.

 **III. Making Judgements about Propitious Industry Structure**

While there are many good reasons for encouraging humility in policy analysts and policy makers, careful analysis and industry observation can provide a basis for making useful judgements about industry structure. The intention is not to forecast the future, even less to provide a comprehensive development plan for the industry. The objective is to identify key economic distinctions that appear to be relatively stable, explore their implications for beneficial industry development, and look for nascent industry trends that may be provide a foundation for promoting such development.

A. Economic Analysis

Connecting end users to a telecommunications network is a local business.[31] Constructing these connections requires careful consideration of local topology and economic geography. Constructing these connections also requires careful consideration of local regulations and politics; wireline network operators need to secure extensive right-of-ways from local governments, and wireless operators need to place antennae. Moreover, in the U.S. in 1990, approximately 28% of all housing units were multiple dwelling units, and the share of such units is significantly higher in other countries.[32] To gain access to end users, network operators often have to enter into highly location-specific, idiosyncratic negotiations with the owners of buildings, campuses, and managed housing tracts. National regulation can play a role in addressing these challenging issues, and the FCC has been actively considering a variety of questions and regulations.[33] Nonetheless, no national regulations are likely to be able to transform end-user connections into a standardized, nationally negotiated and managed service.

---

[31] George Ford, formerly of the Competition Division of the FCC's Office of General Counsel and now at MCI WorldCom, has emphasized this point to me in discussions of industry economics.

In contrast, providing network services is inherently a non-local business relying on standardized routines and infrastructure capabilities. The ubiquity of e-mail services depends on addressing, routing, and formatting standards. Requesting and serving web pages requires additional widely implemented standards. The nature of such standards is largely independent of local knowledge and infrastructure, and the service provided is not related to any geographical location. Customers do not necessarily care where Amazon.com's servers are physically located. For products that it can deliver in electronic form, Amazon.com does not necessarily care where its customers are physically located.[34] Moreover, Amazon.com can expand its capacity to deliver electronic products to customers simply by installing additional standardized hardware.[35] Stock market valuations for companies such as Amazon.com have soared largely because their business models readily scale to global commerce.

Dividing customer value between local connectivity and wide-area network services is a fundamental economic problem. While no amount of head-scratching and eye-gouging can resolve this issue, industry performance will depend heavily on the quality of the arrangements that are worked out. The most important resource for working out such arrangements is relevant information. The best way to generate such information is have customers choose among different combinations of local connectivity and wide-area network services.

B. Institutional Implications

The above economic analysis suggest that good industry performance is likely to depend on the presence of businesses that provide effective separation of local connectivity from wide-area network services. I will refer to businesses that serve this function as service interconnection points or SIPs. SIPs would compete locally in coordinating wide-area network services for local end users. A SIP would lease local facilities providing connectivity to end users and would host and interconnect to facilities distributing wide-area network services. To mediate effectively between local connectivity and wide-area network services, within the relevant geography a SIP should not be owned by either a local facilities provider or wide-area network service provider. This allows a local facilities provider in one area to own a SIP in an area in which the local provider does not provide local facilities. Similarly, a wide-area network service provider could own a SIP as long as that SIP is not connected directly to other SIPs using the wide-area network service provider's facilities.

---

[32] *Promotion of Competitive Networks in Local Telecommunications Markets*, FCC WT Docket 99-217, CC Docket 96-98 (rel. July 7, 1999) para. 29.
[33] Ibid. See also *Telecommunications Inside Wiring*, FCC CS Docket 95-184 (rel. Oct. 17, 1997).
[34] A supporting electronic payment infrastructure, such as that for credit cards, is also necessary.
[35] To the extent that customer support requires human interaction, this is an additional cost factor in scaling service.

Competing, independently owned SIPs would effectively define the product and the value proposition for local facilities builders. The product for local facilities builders would be connectivity from end users to SIPs; thus this connectivity could be defined in terms of the types of attributes currently used to define end-to-end connectivity for leased lines. Competition among SIPs would allow the value of wide-area services to be transmitted to agents considering investments in local facilities. The higher the value to end users of the wide-area services, the greater the amount SIPs would be willing to pay local facilities investors to connect end users to the SIP. By helping to define a local product and value proposition for connectivity, SIPs would foster investment in local facilities.

Enabling localization of investment in communications facilities played a key role in the development of rural telephony in the U.S. Managers of the Bell System, which held Alexander Graham Bell's original telephone patents, believed that telephone service was primarily of value to business users in major cities. In 1894, after seventeen years of commercial activity, the Bell System had installed nearly 90% of its phones for business subscribers.[36] Independent, locally financed commercial telephone companies, community-oriented mutual companies, and farmer cooperatives brought telephony to small agricultural cities and rural areas.[37] By 1920, 38.7% of American farms had telephone service, while only 30% of American households did.[38] Telephone service expanded to cover all of the U.S. under historical conditions that fostered decentralized investment in local access facilities.[39] The geographic structure of local exchanges that was established prior to 1917 still essentially defines the geographic structure for current U.S. interconnection regulation.[40] A set of competing, independently owned SIPs could recreate incentives for decentralized investment in local access facilities.

SIPs would also facilitate low-cost wide-area bandwidth transactions. Some industry participants foresee commodity markets emerging for bandwidth.[41] Such markets could help provide appropriate signals for wide-area network investment and lessen the cost of rolling out new wide-area services.[42] The development of such a market will depend on establishing a widely recognized set of nodes among which bandwidth can be traded. SIPs could serve effectively as nodes for a bandwidth market.

---

[36] Mueller, Milton, *Universal Service; Competition, Interconnection, and Monopoly in the Making of the American Telephone System* (Cambridge, MA: MIT Press, 1997) p. 40.
[37] Ibid, Chapter 6.
[38] Ibid, p. 148.
[39] Ibid, Chapter 12.
[40] In 1917 there were 19,550 local exchange offices in the U.S. Ibid, p. 147. In 1998, local exchange carriers reported a total of 18,700 local exchange switches to the FCC (1998 Statistics of Common Carriers, Table 2.10). The Oct. 1999 LERG lists 22,860 distinct office codes for incumbent local exchange carriers, of which about 4300 appear to be multiple references to listed U S West offices.
[41] Enron has made the most extensive public proposal to data. See www.enron.net/bandwidth/. Companies active in bandwidth brokerage and trading include Arbinet (www.arbinet.com), Band-X (www.band-x.com) and RateXchange (www.ratexchange.com).

A well-developed layer of competing SIPs would provide a lattice upon which new wide-area network-services could be implemented. The largest share of value in wide-area networks is likely to be associated with non-commoditized characteristics such as interconnection services, physical circuit diversity and reliability, and pricing and protocol options.[43] By providing a lattice for implementing such services, SIPs would eliminate the need among competing wide-area networks for a new mode of interconnection in order to provide a new service ubiquitously. Such a lattice would lessen the importance of various forms of peering among wide-area networks and hence decrease industry tensions associated with Internet interconnection.

C. Recent Industry Developments and Institutional Possibilities

SIP-like institutions are already beginning to emerge in the communications industry. One is PAIX, which began operating in 1996 as a center in California for exchanging traffic among ISPs.[44] PAIX states that it is carrier-neutral, not owned by a telco or carrier, and not affiliated with any ISP. PAIX has announced plans to open six additional highly secure facilities for collocation and interconnection among ISPs in the U.S. within a year.[45] Another company offering SIP-like institutions is Equinix, founded in 1998. Equinix builds and operates carrier-neutral and content-provider-neutral facilities it calls "Internet Business Exchanges".[46] Equinix offers for network facilities providers, content providers, and applications service providers a set of buildings with financial grade security, redundant power supplies, private and shared collocation areas, and a wide range of options for interconnecting within the building. Equinix currently has one IBX™ operational in the Washington, D.C. area and recently signed a $1.2 billion contract for constructing more than 30 additional IBX™s in business, financial, and Internet hubs around the world.[47]

---

[42] The development of such a market should not be taken for granted. Attempts to establish commodity markets have historically had a high failure rate even among products with propitious characteristics. See Black, D.G., *Success and Failure of Futures Contracts: Theory and Empirical Evidence* (New York, 1986).
[43] Historically the heterogeneity of user needs spurred the development of private networks. See Gable, David, "Private Telecommunications Networks: An Historical Perspective," in *Public Networks, Public Objectives*, ed. Eli Noam and Aine Nishuilleabhain (Elsevier Science, 1996), pp. 35-49, draft available on the web at www.vii.org/papers/citi509.htm.
[44] All subsequent information about PAIX is from press releases on PAIX's website, www.paix.net. The PAIX center in Palo Alto, CA, was set up by Digital Equipment Corportion. Currently it is a subsidiary of Metropolitan Fiber Network, a seller of dark fiber connectivity.
[45] The locations for the facilities are Tyson's Corner, VA; Atlanta, GA; Dallas, TX; Los Angelos, CA, and an additional facility in Palo Alto, CA.
[46] Unless otherwise noted, the information on Equinix given here is from Equinix's website, www.equinix.com.
[47] According to CNET News, in May 1999, the CEO of Equinix, Al Avery, indicated that Equinix planned to build 15 IBXs domestically. See Heskitt, Ben, "Start-up aims to house Net data exchanges", CNET News.com, May 25, 1999.

Some real estate companies are beginning to provide SIP-like institutions. The Rudin Family, developers and owners of one of New York's largest privately owned commercial and residential real estate portfolios, is a prime example.[48] The Rudin Family developed and owns the New York Information Technology Center at 55 Broad St., Manhattan, which houses a large number of communications and new-media companies. The Rudin Family has established similar facilities at 110 Wall St. and at the former Grumman Aircraft factory on Long Island. The Rudin Family recently bought from AT&T a former AT&T long lines switching center at 32 Avenue of the Americas. The building will be renovated and called the New York Global Connectivity Center. It will house network transport providers, web-hosting companies, Internet companies, and switch companies, and it will provide extensive support for in-building interconnectivity.

Most of the SIP-like institutions described above are located in major cities and do not provide services to end users. In residential and rural areas, ISPs moving to offer their end users a variety of network services may evolve into SIPs. Dial-up Internet connectivity has become for ISPs a low-margin, commodity service that cannot sustain their businesses. ISPs are thus seeking to develop value-added businesses such as web hosting, video-conferencing, e-commerce, and variety of other wide-area network services. An impediment to ISPs ability to offer their customers new services is the lack of competition in local telephony in residential and rural areas. However, as ISPs assemble increasingly appealing offerings of network services, they will generate strong incentives for the entry of local facilities providers who can connect end users to these services. Given that numerous ISPs provide local service in almost all regions of the U.S.[49], they could be important to the development of competing, independently owned SIPs that cover all of the U.S.

## IV. A Feasible Policy Lever for Improving Industry Structure

Changes in voice telephony regulation to promote the development of SIPs could help overcome the structural weaknesses that are appearing in Internet and telephony competition. In particular, voice telephony regulation could seek to establish a geographically comprehensive lattice of competing, independently owned certified SIPs. Becoming a certified SIP would involve gaining a privileged position for voice telephony call termination in exchange for adhering to certain ownership restrictions. All telephony service providers in defined SIP regions would be required by regulation to provide zero-price call (circuit-switched voice, fax, and dial-up modem) termination for calls delivered to chosen certified SIPs in the SIP region associated with the called customer. The owner of a certified SIP would not be allowed to own facilities for local connectivity in the area in which the certified SIP is located. A certified SIP would also not be allowed to own network facilities connecting to other certified SIPs.

---

[48] The subsequent information is from press releases at www.55broadst.com and Branson, Ken, "AT&T Sells Former Long Lines Building for Telco Hotel", *Phone+,* 12/16/1999 (www.phoneplusmag.com). SIP-like institutions are generally called "telco hotels" in the trade press. For discussions of telco hotels, see Branson, Ken, "No Vacancy, Telco Hotels Can't Go Up Fast Enough", *X-Change Magazine*, 4/1999 (www.x-changemag.com); Marshall, Jonathan, "Telco Hotels Fill Up Fast", *San Francisco Chronicle*, July 2, 1998, reprinted on the web at www.boradlink.com/press/colomotion.html. Other non-telecom companies that offer or own colocation services include Colomotion (www.colomotion.com), Hudson Telegraph Associates, Switch and Data Facilities Corp. (www.switchfacilities.com), Taconic Investors, Telecom Real Estate Service, and Telehouse (www.telehouse.com).

[49] See Greenstein, Shane, "Universal Service in the Digital Age: The Commercialization and Geogrpaphy of U.S. Internet Access," NBER Working Paper No. W6453 (March 1998), available at www.nber.org.

Further necessary decisions about certified SIPs would depend on institutional circumstances. One issue is the geographic areas associated with SIPs' voice termination roles and transport facility ownership restrictions. In the U.S., a natural choice is LATAs. Each state regulatory commission might certify, for a fixed term of 5 years, 3-5 independently owned SIPs in each LATA in the state, with all local telephony operators in the state having responsibility to terminate calls from at least two of those SIPs. Since there are 236 LATAs covering the U.S., such a program would lead to roughly 750-1000 certified SIPs spread throughout the U.S. If all voice traffic, including local calls, passed through these SIPs, they would have to support 1.9-2.5 Gbps of voice traffic.[50] This is about the volume of peak data bandwidth through a major U.S. Internet interconnection point in late October, 1999.[51]

There is a range of institutional possibilities for facilities and ownership of SIPs. Independently owned Internet or private network interconnection points might be candidates to be certified SIPs. Highly capable ISPs meeting the transport facility ownership restrictions might also be candidates to be certified SIPs. National network operators, many of whom are building large data centers, might be willing to divest transport facilities to some data centers in order to make them candidates to be certified SIPs.[52] In addition, regulators or antitrust authorities could consider requiring large incumbent LECs to divest some tandem switching offices so as to create an interconnection structure more conducive to controlling incumbent LEC market power.[53]

A. Considerations of policy feasibility

Changing voice telephony regulation is a much more propitious policy direction for influencing evolving industry structure than is establishing new regulations for Internet peering. Large incumbent LECs are widely recognized to have market power in providing local telephony. In contrast, market structures for Internet services are highly dynamic, and market power arguments with respect to Internet services typically depend significantly on speculation about future developments. Moreover, while there is widespread, deeply rooted hostility toward changing the regulatory framework for Internet services, telephony regulation over time has gone through a series a major new regulatory initiatives. Associated with that history is the FCC's extensive knowledge and experience with implementing telephony regulation. There is no similar knowledge and experience with respect to regulating Internet interconnection. In addition, there is significant dissatisfaction with the current state of telephony regulation, and the Telecommunications Act of 1996 gives the FCC broad forbearance authority with respect to almost all of its regulations. Changes in voice telephony regulation that promote SIPs could be accompanied with a dramatic reduction in a wide range of other regulations that would no longer be part of this new implementation of a pro-competitive, deregulatory national policy framework for the industry.

---

[50] Calculation based on total voice bandwidth given in Table 1. Voice calls among incumbent telephony customers would not necessarily pass across SIPs.
[51] Based on bandwidth for MAE East. See 208.234.102.97/MAE/east.aggr.overlay.html.
[52] AT&T, UUNet, PSINet, Qwest, and Intel are talking about building about 25 new data centers each this year. See Gerwig, Kate, "Salving Future Services", *tele.com*, 1/10/2000 (at www.teledotcom). Level 3 has also built more than 25 data centers that offer a wide range of services to collocating customers.
[53] Large incumbent LECs might, in fact, find it advantageous, from the perspective of transforming industry structure and maximizing asset value, to do this.

While data traffic is growing much more rapidly than voice traffic, the value and magnitude of voice traffic is still sufficient to influence strongly the over-all structure of network interconnection. The first data column of Table 1 shows total RBOC inter-office bandwidth in use. Subsequent data columns show the total bandwidth required for RBOC originated voice calls in different categories. Assuming that all local voice traffic travels between RBOC local exchanges, the total bandwidth of RBOC interoffice facilities in use for non-voice services in 1998 was 2.4 times greater than bandwidth needed for voice services. Most of the non-voice bandwidth is for leased line services, whose bandwidth has been growing about 40% per year since 1989.[54] Internet bandwidth in mid-1998 was probably about 110Gbps and it is growing about 100% per year.[55] The most important point to take from Table 1 is that voice services still account for an important share of network bandwidth, although within a few years voice bandwidth will be insignificant.[56] This means that becoming a distinguished interconnection point for voice telephone can play an important role in giving an interconnection point industry salience.

---

[54] Despite the significant growth of private networks, there has been relatively little analysis of them. Dunn, Donald A. and M. Gens Johnson, "Demand for Data Communication," *IEEE Network* (May, 1989), p. 8-12, provides an informed perspective late in the 1980's. Dunn and Johnson foresaw the wide-area interconnection of computer networks that created the Internet. They estimated that data revenue was growing 23% per year in 1988, and anticipated that data revenue would account for more than half of common carrier revenues in 1997. In 1997, according to the FCC *SOCC* Table 2.9, data revenue accounted for about 6% of local exchange carriers common carrier revenue. Based on data in AT&T's 1998 Annual Report, I estimate that about 20% of AT&T's revenue is data revenue.

[55] Coffman, K.G, and A.M. Odlyzko, "The size and growth rate of the Internet," (available on the web at www.research.att.com/~amo) estimated the effective bandwidth of the Internet core at 75Gbps at year-end 1997. They also estimate that, with the exception of a spurt in 1995 and 1996, the trend growth rate of core Internet bandwidth is 100% per year. Coffman and Odlyzko estimate U.S. long distance voice bandwidth at 350Gbps at year-end 1997. Since RBOCs account for about 70% of U.S. local access lines, the Coffman-Odlyzko long distance voice bandwidth figure is roughly in accord with the interLATA toll figures in Table 1 above. On the other hand, Coffman and Odlyzko's figure for private line and public data networks, 370Gpbs at year-end 1997, appears to be a significant underestimate.

[56] Forbes.com reported that Bing Yang, the chief technology officer and cofounder of Convergent Networks, which focuses on voice networks, stated that AT&T's network carries 850 terabytes of voice traffic per day as compared to 33 terabytes of data traffic. See Malik, Om, "Telecom titans", *Forbes.com*, 9/8/99 (www.forbes.com/tool/html/99/sep/0908/feat.htm). The specific nature of these measurements is unclear. Nonetheless, it is worth noting that utilization rates for long distance switched voice circuits (33%) are almost an order of magnitude greater than utilization rates for data circuits (3-5%). See Odlyzko, Andrew, "The Internet and other networks: Utilization rates and their implications," (available on the web at ww.research.att.com/~amo) AT&T Labs – Research, Sept. 12, 1998. The inter-office bandwidth data in Table 1 refer to the bandwidth of circuits in use, not the volume of traffic passing through those circuits.

| | interoffice bandwidth | interLATA toll | intraLATA toll | local | total voice |
|---|---|---|---|---|---|
| | \multicolumn{5}{c}{Table 1 RBOC Bandwidth in Use (in Gbps)} | | | | | |
| 1998 | 6,291 | 217 | 54 | 1,602 | 1,874 |
| 1997 | 4,128 | 221 | 61 | 1,540 | 1,822 |
| 1996 | 3,376 | 206 | 58 | 1,501 | 1,766 |
| 1995 | 2,762 | 194 | 66 | 1,456 | 1,716 |
| 1994 | 1,654 | 172 | 65 | 1,396 | 1,633 |
| 1993 | 1,304 | 156 | 64 | 1,340 | 1,560 |
| 1992 | 934 | 142 | 64 | 1,299 | 1,505 |
| 1991 | 729 | 135 | 66 | 1,256 | 1,457 |
| 1990 | 493 | 133 | 56 | 1,217 | 1,407 |
| 1989 | 346 | 157 | 55 | 1,174 | 1,386 |

Notes: Interoffice bandwidth calculated based on data in RBOC price cap annual filings. Telephony bandwidth based on FCC SOCC call volumes, estimated call times, and 9000 minutes/month/per 64 Kbps circuit.

B. Some consideration of costs and benefits[57]

A requirement that all local telephony providers in a defined geographic area provide zero-price call termination from at least two certified SIPs in the area has relatively small costs and large benefits. This requirement promotes the concentration of network traffic at certified SIPs and thus helps to promote SIPs' industry significance in future network development. This requirement also provides an administratively simple telephony interconnection regime that would allow network operators to provide flat-rated telephone service. It would eliminate huge battles such as those that have occurred in the U.S. over reciprocal compensation for switched circuit minutes associated with dial-up Internet connections.

Competition in local telephony by itself will shift more voice traffic into inter-office networks. Traditionally, local exchange offices were designed around local calling communities so that local calls could be completed without the need for inter-office transport. When neighbors are connected to competing local telephone companies, local calls require inter-office transport.[58] As the figures in Table 1 suggest, because local call volumes are high relative to intraLATA and interLATA toll calling, competitors shifting even a small share of an incumbent LEC's local call bandwidth to inter-office transport can result in a large percentage increase in inter-office voice transport.[59]

---

[57] For further discussion of questions and objections raised regarding this proposal, see Galbi, Douglas, "Transforming Network Interconnection and Transport: Policy Direction Summary," currently available on the Social Science Research Network (www.ssrn.com), forthcoming in *info*.
[58] To lower their customers' phone bills for dial-up Internet access, some ISPs have sought to become "virtual neighbors" of their customers. They do this by acquiring a number for each local calling area for
19

In a competitive industry, reducing interconnection management costs is probably more important than reducing the demand for inter-office voice transport. Given the magnitude of total inter-office bandwidth, doubling or tripling the amount of inter-office voice transport would not require a major re-dimensioning of the over-all network. On the other hand, managing interconnection involves exchanging traffic predictions at each interconnection point and coordinating the installation and maintenance of new interfacing bandwidth. Such processes are administratively complex, error-prone, and not subject to rapid technological improvements like those driving down bandwidth and switching costs. Nonetheless, industry experience thus far shows incumbents often seeking to require competitors to interconnect with them at a relatively large number of local offices.[60]

New local telephony providers terminate calls to their customers from relatively few publicly advertised offices. Consider Table 2, which documents some aspects of telephony network structure in the greater New York City metro area (LATA 132). The first data column in Table 2 gives the number of rate centers served. Rate centers are a historically determined geographic partition of an area: the number of rate centers served is a rough index of the scope of a telephony provider's coverage. The second data column of Table 2 shows offices advertised in the Local Exchange Routing Guide (LERG) as delivery points for calls to the provider's customers.[61] New local telephony providers cover a significantly larger number of rate centers per call delivery point than does Bell Atlantic. This suggests that new local telephony providers are not seeking to economize on transport costs by having telephony providers deliver calls to them close to their end customers.

The time and cost of establishing points of presence is not hindering the ability of new local telephony providers to establish more termination points for calls to their customers. Through July 1999, Bell Atlantic-NY had provided 750 physical collocation

---

which they provide dial-up Internet access and having their communications provider terminate all these numbers to the same physical point. One result is that the incumbent operator sees more local call minutes traveling through its interconnection trunks.

[59] For a discussion of this problem in Austria, see Merka, Martin, Manfred Nussbaumer, and Ernst-Olay Ruhle, "The Influence of Interconnection Demand on Traffic Flows and Network Design for an Incumbent Operator – The Austrian Example," paper presented at the 17'th Annual ICFC Conference, Denver USA, June 16, 1999.

[60] Bell Atlantic-New York has asked the NY Public Service Commission (NYPSC) to require that interconnecting LECs establish a geographically relevant interconnection point (GRIP) in every rate center that the LEC serves, unless the interconnecting carriers negotiate alternative arrangements. See NYPSC, Case 99-C-0529, Opinion No. 99-10 (Aug. 26, 1999), p. 48, 62. The NYPSC rejected the GRIP proposal.

[61] The LERG is produced by the Traffic Routing Administration (TRA), Telcordia Technologies. The FCC has published LERG data on carriers' counts of NXX to provide an indication of the development of competition. See *Local Competition: August 1999*, Industry Analysis Division, Common Carrier Bureau, FCC, available on-line at http://www.fcc.gov/Bureaus/Common_Carrier/Reports/FCC-State_Link/fcc-link.html.

arrangements in 175 central offices.[62] Of these collocation arrangements, 137 are in the greater New York City area (LATA 132).[63] Nonetheless, as Table 2 shows, new telephony providers are using only 13 collocations in BA offices in LATA 132 as points for collecting calls from other networks. When such arrangements are used, the larger new local telephony providers use exclusively BA tandem offices. This fact further suggests that the geography of interconnection points for terminating voice telephony has thus far been determined by the historical location of a relatively few, large incumbent offices. Policy that requires voice telephony call termination from certified SIPs represents a dramatic change only in the sense that it shifts voice telephony interconnection to a non-adversarial environment, i.e. competing, independently owned SIPs.

| Table 2 Voice Telephony Delivery Points in LATA 132 | | | | | |
|---|---|---|---|---|---|
| Local Telephony Provider | Rate Centers (a) | Delivery Points (b) | Ratio (a)/(b) | Colo's in BA Offices | Colo's in BA Tandems |
| Bell Atlantic (wireline) | 126 | 167 | 0.8 | | |
| AT&T (wireline) | 50 | 13 | 3.8 | 3 | 3 |
| MCI WorldCom | 32 | 7 | 4.6 | 0 | 0 |
| Nextlink | 28 | 1 | 28.0 | 0 | 0 |
| Allegience | 23 | 5 | 4.6 | 3 | 3 |
| Cablevision Lightpath | 20 | 7 | 2.9 | 0 | 0 |
| RCN | 20 | 6 | 3.3 | 4 | 4 |
| American Network, Inc. | 18 | 1 | 18.0 | 0 | 0 |
| Frontier | 18 | 5 | 3.6 | 4 | 4 |
| WinStar | 15 | 1 | 15.0 | 0 | 0 |
| Level 3 Comm. | 15 | 2 | 7.5 | 0 | 0 |
| | | | | | |
| all other than BA | 83 | 77 | | 13 | 6 |
| Note: Based on LERG data current for 10/1/99. AT&T (wireline) consolidates entries for ACC National Telecom, AT&T Local, and Teleport. MCI WorldCom considates entries for Brooks Fiber, MCIMetro, and WorldCom. | | | | | |

**V. Conclusion**

Policy analysts and policy makers should consider the merits of different competitive structures in the telecommunications industry. Significant weaknesses in industry structure are apparent in the Internet and in the development of local telephony competition. Pro-competitive regulation for voice telephony is not adequately considering over-all industry structure; instead, it appears to be largely driven by particular, narrow requests for pro-competitive interventions. Nonetheless, voice telephony regulation will have an enduring effect on industry structure even when voice telephony is a relatively unimportant network service. Armed with an understanding of the challenges confronting both Internet services and voice telephony, such a legacy can become a tool for improving industry performance.

---

[62] *Bell Atlantic-NY Section 271 Application*, Lacouture/Troy Declaration, para. 29.
[63] Calculated based on LERG data on LATA's and wirecenters and BA's list of offices where collocation has been provided. This list is available on the web at http://www.bellatlantic.com/tis/bacolloc.htm.